\documentclass[12pt]{article}

\usepackage{amsmath,amssymb,amsfonts,amsthm}
\usepackage{bbm} 
\usepackage{graphicx}
\usepackage{hyperref}
\usepackage[numbers,sort&compress]{natbib}
\usepackage{geometry}
\usepackage{enotez}
\setenotez{list-name=Endnotes} 

\title{Two Shades of Quark Color: Parallel Canons across the Cold War Divide}
\author{Vitaly Pronskikh\thanks{The Center for Philosophy of Science, University of Pittsburgh, email: \texttt{vpronskikh@gmail.com}}}
\date{\today}

\begin{document}

\maketitle

\begin{abstract}
The introduction of the color quantum number is conventionally narrated as a linear progression from the quark-model statistics paradox to quantum chromodynamics (QCD). This paper challenges that teleology by arguing that ``color'' emerged as two conceptually distinct constructs during the Cold War. The first, originating with Han and Nambu and culminating in QCD, conceived of color as a local gauge charge, the source of a fundamental force mediated by gluons. The second, developed at the Joint Institute for Nuclear Research (JINR) in Dubna, treated color as a hidden, three-valued label—a statistical and structural property within a composite, S-matrix-inflected hadron model. We trace these parallel narratives, linking the Dubna approach to a holist epistemology that prioritizes observable amplitudes and global constraints, and the QCD approach to a reductionist program grounded in micro-dynamics. A case study of Fermilab’s E–36 experimental chain (1970–78) shows how an observables-first design—tuned to S-matrix and Regge restrictions on forward elastic scattering—performed robustly within its natural domain but was ultimately discontinued amid declining theoretical interest and involvement. The subsequent hegemony of QCD retroactively projected its gauge-theoretic conception of color onto history, erasing this epistemic diversity. We conclude that the marginalization of Dubna's structural color was not merely a political outcome of the Cold War but a result of deep ontological and philosophical divergences, advocating for a domain-sensitive pluralism in the historiography of particle physics.
\end{abstract}

\section{Introduction}
The introduction of the \textit{color} quantum number is often taken—retrospectively—as the hinge between
phenomenological schemes for organizing hadrons and a dynamical field theory of their constituents,
quantum chromodynamics (QCD) \cite{Fritzsch1973}. The canonical storyline runs from the quark model’s
statistics paradoxes \cite{GellMann1964,Zweig1964} through the postulation of an additional internal
degree of freedom \cite{HanNambu1965,BST1965,Struminsky1965} to the maturation of a non-
Abelian gauge account in which color is a local charge
\cite{tHooft1971,tHooftVeltman1972,Fritzsch1973,GrossWilczek1973,Politzer1973}. This paper resists that
teleology. We argue that “color” emerged as \emph{two} conceptually distinct constructions: a hidden,
three-valued \emph{label} introduced within composite-hadron phenomenology (the Dubna line), and a
\emph{local gauge charge} in Yang–Mills theory (the Han–Nambu thread and the later QCD synthesis). These
constructions grew in partially overlapping but differently normed research cultures, partially incommensurable and yielding parallel
—sometimes rival—epistemic orientations rather than a single, linear path to QCD.

The foundational work of Matveev and Tavkhelidze, as detailed in their retrospective \cite{Matveev2006}, 
reveals a conceptual lineage originating from the JINR
\cite{BST1965,Struminsky1965}. Here, the impetus for color arose from a deeply \textit{holist}
and compositeness-oriented program \cite{BST1965}. The primary challenge was to construct a
relativistically invariant dynamical model of hadrons in which quarks, though not observed in isolation, could
be treated as “real” physical objects. The statistics problem,
arising from the symmetric ground-state wavefunctions of baryons, was resolved by introducing a new quantum
number so that the total wavefunction becomes antisymmetric \cite{Struminsky1965,BST1965}. In this
framework, color was fundamentally a \textit{statistical and structural} property, a necessary attribute to
ensure the consistency of a composite model in which the whole (the hadron) was ontologically prior to its
parts (the quarks). The subsequent development of the “quasi-free quark model” and the derivation of scaling
laws from the principle of self-similarity \cite{MMT1972LNC}
further cemented this approach within an S-matrix and bootstrap-inspired tradition that privileges asymptotic
behavior and global consistency conditions over explicit local gauge fields
\cite{Eden1966,Chew1962,Collins1977}.

Conversely, the approach pioneered by Han and Nambu \cite{HanNambu1965} and later consolidated into QCD was 
explicitly \textit{reductionist}. Its aim was to identify the fundamental fields and their symmetries. In this
paradigm, color was introduced \textit{ab initio} as a \emph{local} $SU(3)$ symmetry of quark fields with
eight gauge bosons mediating the interaction \cite{HanNambu1965,YangMills1954}. The focus was not merely on curing
the quark-model statistics paradox but on constructing a Lagrangian field theory based on the local gauge
principle. Quarks were the fundamental entities, and their interactions, mediated by gauge bosons (gluons),
were the primary objects of study. The mathematical viability of such a non-Abelian gauge theory depended on
proofs of renormalizability \cite{tHooft1971,tHooftVeltman1972}; the specific assignment of color as the
charge of the strong force was articulated in the color-octet gluon picture \cite{Fritzsch1973}. 
With the nearly simultaneous discovery of asymptotic freedom~\cite{GrossWilczek1973,Politzer1973} and Wilson's non-perturbative lattice formulation~\cite{Wilson1974}, confinement could be treated as a dynamical property (e.g., Wilson’s area law on the
lattice) rather than a pre-theoretical axiom. Asymptotic freedom provided the high-energy handle, while the Wilson loop and lattice QCD offered a framework to understand low-energy confinement.

The subsequent and overwhelming success of QCD has retroactively projected its 
gauge-theoretic conception of color onto the entire history of the
idea, effectively erasing the epistemic diversity of the Cold War physics landscape.
The distinct philosophical underpinnings—the holist compositeness of Dubna versus
the reductionist gauge theory —have been collapsed into a single,
streamlined narrative.

This paper reconstructs these two parallel narratives and clarifies their association with holism
and reductionism, respectively. We trace their historical trajectories and situate them within the
larger physics programs and their cultural–institutional contexts. To ground the analysis, we
present a case study of the Fermilab E–36 experimental chain (1970–1978), designed and interpreted
within an S-matrix/Regge framework and aligned with Dubna’s automodelity and quark-counting
program, thereby showing how an observables-first, amplitude-based idiom shaped experimental
design, analysis, and success criteria. 

We conclude by reassessing Dubna’s contributions within the broader epistemic shift that bifurcated the strong-interaction canon into holist S-matrix and reductionist QCD strands. This bifurcation, more than institutional politics, helps explain the muted place of Dubna’s hidden-index conception of color in the contemporary QCD canon, and we close by sketching the implications for how paradigm change reshapes historical memory in high-energy physics.

\section{Two Colors — Two Narratives}
\label{sec:two-colors}
The resolution of the quark statistics paradox in the mid-1960s emerged not as a singular discovery but as parallel developments within distinct theoretical frameworks. The Dubna school and Western approaches conceived of ``color'' with fundamentally different physical interpretations and mathematical implementations, reflecting their divergent research traditions.

\subsection{Dubna’s ``color as label'' (1965)}
In early 1965, Dubna authors (see Endnote~{\endnote{\label{struminsky1} I am aware of the controversy over Struminsky’s primacy in the Dubna quark-color proposal, particularly the exchange between \cite{Struminsky1965} and \cite{Petrov2009}. I do not adjudicate that dispute here. For purposes of textual citation, I rely on \cite{Struminsky1965} as the only available English translation of Struminsky’s preprint; accordingly, I sometimes refer to the Dubna quark-color model as “Struminsky’s work,” insofar as his JINR preprint is the first published record of the result. At the same time, the Bogolyubov group’s working style was manifestly collectivist; irrespective of who first formulated particular claims, the model can be read as the work of a collective subject responsible for the idea and its development. This stance parallels the “collective ontologies” that characterize experimental collaborations \cite{Pronskikh2019EPS}. In this spirit, individual names in what follows function as convenient signifiers of strands within a shared research program rather than as exclusive claims of authorship.}) proposed an \emph{additional internal quantum number} for quarks—what would
soon be called \emph{color}—precisely to reconcile the SU(6) quark spectroscopy with Fermi--Dirac/Pauli
constraints in the composite hadron model frame \cite{BST1965,Struminsky1965,Tavkhelidze1965,Matveev2006}.
In this narrative, color first functions as a \emph{hidden index} that antisymmetrizes the otherwise
symmetric spin–flavor wavefunctions and enforces selection rules at the level of residues in amplitudes,
while hadrons observed in nature are taken to be \emph{color neutral} (singlets)
\cite{Matveev2006,MatveevBaldin2025}. 
%
The JINR work of 1965 develops a \emph{dynamic} composite-hadron model in which quarks are degrees of freedom yet permanently confined, and only color-singlet
hadrons appear as asymptotic states (the free, experimentally detectable particles (like hadrons) long before or after an interaction, which form the basis of the S-matrix formalism)~\cite{MatveevBaldin2025}. Color here \emph{labels} unitary-equivalent
internal states that do bookkeeping in the residue algebra, the mathematical structure governing the strengths (residues) of bound-state poles in scattering amplitudes, encoding selection rules and symmetry requirements. No explicit gluon dynamics is posited at this
stage.

Historically, Struminsky’s 1965 Dubna work (see Endnote~\endnote{\label{struminsky2} Boris Vladimirovich Struminsky (1939–2003) was a Soviet, Russian, and Ukrainian theoretical physicist. From 1962 to 1965, he was a doctoral student at the Steklov Institute of Mathematics in Moscow, where he continued as a junior researcher in 1965–1966. From 1966 to 1971, Struminsky worked as a researcher at the Joint Institute for Nuclear Research (JINR) in Dubna. Beginning in 1971, he held research positions at the Institute for Theoretical Physics of the Academy of Sciences of the Ukrainian SSR—now the Bogolyubov Institute for Theoretical Physics of the National Academy of Sciences of Ukraine—in Kyiv.~\cite{StruminskyWiki}.}) introduced an \emph{additional internal
quantum number} for quarks to reconcile SU(6) with Fermi–Dirac statistics; it did
\emph{not} posit an SU(3) local gauge theory or an octet of gluons
\cite{Struminsky1965}. The first move to put the new degree of freedom on an SU(3)
gauge footing—explicitly with \emph{eight} vector fields as interaction carriers—
was by Han and Nambu in 1965 \cite{HanNambu1965}; see also Matveev–Tavkhelidze’s 
retrospective noting Nambu’s eight vectors as the gluon prototype
\cite{Matveev2006}. The parafermi statistics route, introduced by Greenberg\cite{Greenberg1964}, provided a global, hidden degree of freedom. While it could later be connected to a gauge theory—for instance, via the Green ansatz leading to an SO(3) model with three gauge bosons, which is phenomenologically distinct from QCD—its original formulation was not that of a local gauge theory.
Despite Struminsky being first to use an SU(3) symmetry in color studies, in his formulation it served as a \emph{global} SU(3)–type label (three unitary-equivalent quark states) rather than a local gauge symmetry \cite{Struminsky1965,Matveev2006}. The parafermi route was later shown to be equivalent to the SU(3) color model in its classification of the allowed color-singlet hadronic states, not as a dynamically equivalent SO(3) gauge theory~\cite{GreenbergZwanziger1966,Greenberg2008}.

The Dubna program subsequently developed the "quasi-free quark model" where color neutrality ($\epsilon^{\alpha\beta\gamma}$ for baryons, $\delta_\alpha^\beta$ for mesons) defined asymptotic states while quarks remained permanently confined through dynamical mechanisms external to the color concept itself \cite[pp. 308–309]{Matveev2006}.

%

\subsection{Parallel fixes and divergences (1964–66).}
\textbf{Greenberg (1964):} parafermi statistics of order~3 provides a \emph{global, hidden} degree of
freedom allowing three identical quarks in a symmetric spin–flavor state without violating Pauli, thereby
solving baryon spectroscopy kinematics without introducing a local SU(3) gauge
\cite{Greenberg1964,Greenberg2011,Greenberg2008}. Dubna accounts emphasized, however, that \emph{parastatistics is not equivalent to color SU(3)} as a gauge symmetry and generally yields different
physical consequences \cite{Matveev2006}.%
%

\noindent \textbf{Han–Nambu (1965):} a \emph{double SU(3)} scheme with explicit \emph{vector gauge fields}
(proto-gluons) put color on a local footing, even though their model adopted \emph{integer quark charges}
(not part of QCD’s final architecture) \cite{HanNambu1965,Greenberg2008}. This was the first
step from “color as hidden residue index” toward \emph{color as an acting charge} with gauge carriers.%
%

\noindent \textbf{Early colorless-hadron hypothesis} assumed that only singlets appear as asymptotic states
(“confinement” as a working premise long before lattice proof); this sits naturally with the Dubna residue-algebra view \cite{Matveev2006,MatveevBaldin2025}.%

These parallel developments reveal the period's conceptual fluidity: identical phenomenological requirements (statistics, confinement) admitted multiple physically distinct solutions, with the choice between them reflecting deeper commitments to either composite models or emergent gauge theories.
%

\subsection{Consolidation into QCD (1971–79).}
The convergence toward modern QCD occurred through independent theoretical developments that transformed color from a classification device to a dynamical charge. The proof by 't Hooft and Veltman \cite{tHooftVeltman1972} that non-Abelian gauge theories are renormalizable provided the essential mathematical foundation for treating color as a local symmetry.

Renormalizable non-Abelian gauge theory provided the viability proof for local color dynamics (’t~Hooft–
Veltman program building on Faddeev–Popov gauge fixing); asymptotic freedom then made the strong coupling
run the \emph{right} way \cite{tHooftVeltman1972,FaddeevPopov1967,GrossWilczek1973,Politzer1973}. 
On the theoretical side, RG-improved operator-based methods—epitomized by DGLAP (Dokshitzer–Gribov–Lipatov–Altarelli–Parisi equation) evolution of parton distributions — supplanted earlier S-matrix approaches by enabling calculable links between short-distance dynamics and inclusive cross sections.~\cite{GribovLipatov1972,Dokshitzer1977,AltarelliParisi1977}. In the same
reductionist program, three-jet topologies at ~\cite{Brandelik1979} and the subsequent event-shape/4-jet-angle/multiplicity analyses were \emph{interpreted within the QCD
framework} (parton showers + hadronization models) as signatures of hard non-Abelian radiation; the extraction of the color-factor ratio \(C_A/C_F\) is thus a
\emph{model-mediated} inference, with
``color'' functioning as the coupling label of the gauge sector, not as a directly
observed S-matrix quantum number \cite{TASSO1979,Brandelik1980,PDGQCD2020}.
%

\subsection{The bridges (micro \texorpdfstring{$\leftrightarrow$}{↔} macro).}
By “bridges” I mean not new forces but \emph{translation rules} between two descriptive layers of the same process. They do not show us quarks or gluons “in the data”; rather, they license moving from amplitude language (poles, cuts, residues) to averaged, colorless spectra in a controlled way.
Firstly, dualities in theory function as bilingual dictionaries. Finite–energy duality (Dolen–Horn–Schmid) and Bloom–Gilman duality say, in effect, that when we average over many resonances, the hadronic story and the short-distance scaling story can be written as two dialects of one account \cite{DHS1967,BloomGilman1970}. This is a \emph{consistency norm}, not a literal ‘microscope’ on quark dynamics, but rather a rule for translating residue sums into scaling integrals.

Secondly, sum rules function as normative criteria. Current algebra with dispersive integrals (Adler; Adler–Weisberger) ties symmetry commitments to measurable rates \cite{AdlerWeisberger1965}. Philosophically, these are not discoveries of hidden objects but \emph{obligations} our amplitude calculus must satisfy if the symmetry talk is to earn its keep.
The Shifman–Vainshtein–Zakharov SVZ (ITEP, Moscow) sum rules played a role of an accountability ledger. The SVZ program (OPE + dispersion) relates “micro” coefficients (condensates in operator products) to “macro” spectral weights of hadrons \cite{SVZ1979}. Color remains absent from the S-matrix spectrum; what is present is a disciplined matching between two descriptions—one written in operator coefficients, the other in colorless spectral functions.

These bridges keep Dubna’s residue algebra and QCD’s reductionist calculus in dialogue. They let a hidden index live in the residue bookkeeping while the gauge–theoretic color only \emph{acts} inside the calculation. 
%

\subsection{What's at Stake Conceptually}
\label{subsec:conceptual_stakes}

The essential distinction between the two color concepts can be captured aphoristically: \textit{Dubna ``color'' = symmetry of amplitude/residue; QCD color = source of the force}. 
In the 1965 Dubna scheme, quarks are treated as confined \emph{constituents} (entity-level commitment), while “color” functions as a \emph{global} SU(3)-type index that stabilizes the residue algebra (Pauli/SU(6) consistency and selection rules). Thus color is real \emph{as a structural condition}, not as a causally efficacious charge. The subsequent QCD consolidation \emph{refunctionalizes} the same SU(3) into a \emph{local} gauge symmetry: color becomes a dynamical charge (with gluon carriers), i.e., real \emph{as a source term} in the field equations. In short: quarks-as-entities with color-as-constraint (Dubna), versus quarks-as-entities with color-as-cause (QCD).

In the Dubna framework, color governed the combinatorial structure of hadronic wavefunctions and selection rules—a property of composite systems rather than fundamental fields. As Matveev emphasizes, their approach treated "colored quarks as real physical objects defining the hadron structure" within a dynamical model where confinement was presupposed rather than explained \cite[pp. 307–308]{Matveev2006}.

In QCD, color became the charge sourcing the gluon field—a fundamental property requiring explanation through dynamics (asymptotic freedom, confinement). The coexistence of these interpretations was mathematically possible precisely because color never appears as an asymptotic quantum number—the "hidden index" could function equally well in residue algebra or as the source of a confining force.

This distinction maps onto a broader epistemic divide: compositeness models treat properties as emergent from symmetry and structure, while gauge theories derive them from localized interaction mechanisms encoded in field dynamics.
Struminsky models baryons as \emph{composites of three quarks} and then adds an \emph{additional quantum number} so that the residue algebra respects Pauli/SU(6) without ever populating the asymptotic spectrum with colored states \cite{Struminsky1965,BST1965}. In short, quarks are the \emph{constituent variables} of a composite description, while color is a \emph{global SU(3)-type index} that makes the whole amplitude narrative cohere—precisely a holist use of structure, not yet endowed with genuine dynamics. While not derived from S-matrix theory, Struminsky’s approach reflects its holist spirit: it privileges observable, colorless states (only colorless hadrons appear as $S$-matrix states); treats color as a structural criterion rather than a gauge-mediated force; stabilizes the amplitude bookkeeping rather than a local cause; emphasizes internal consistency over microscopic causation—placing it epistemologically closer to compositional holism than to dynamical field theory.

\section{Holism and reductionism in particle physics}

In physics, ``reductionism'' is not the naive claim that the world is nothing but tiny billiard balls; it is the working program of \emph{explaining regularities by a small set of micro-entities and interaction laws} and then tracking how those laws \emph{flow with scale}. The lineage runs from Greek atomism (Leucippus, Democritus, Epicurus) through the early-modern mechanical philosophy (Descartes, Boyle) to the quantum-field-theoretic version in which \emph{local fields and gauge symmetries} carry the micro-story \cite{KRS1983,Lucretius,Descartes1644,Boyle1661}. Wilson’s RG reconciles reduction and emergence. It allows for derivation across scales, but recognizes that macroscopic structures exhibit autonomy: they obey their own effective laws, shaped but not rigidly dictated by microscopic inputs. It is a holist-compatible form of scale-sensitive reductionism~\cite{WilsonKogut1974}. In this refined sense, reductionism means (i) positing micro-dynamics and (ii) computing how those dynamics organize phenomena across scales via RG and effective field theory (EFT) \cite{Weinberg1979}. Wilson’s renormalization group does double duty: it shows why macroscopic behavior can be \emph{insensitive} to microscopic minutiae (universality, fixed points)—a holist moral—while also supplying the \emph{method} for deriving scale–dependent laws from a local microdynamics (decoupling, EFT), i.e., a workable, non-naive reductionism \cite{WilsonKogut1974,Weinberg1979}. \cite{CastellaniMargoni2022} review debates on RG explanations—especially Morrison’s critique—and argue that while these explanations involve crucial mathematical structures, their power “cannot reside exclusively in [their] mathematical character.” In particular, RG does not yield a purely formal account; it requires a physical interpretation of scale transformations to function explanatorily. This underscores that even in reductionist frameworks, explanatory force often depends on embedding abstract tools within a physically meaningful narrative.

For the strong interaction, the reductionist bet is that a \emph{local} SU(3) gauge theory with quarks and gluons suffices to organize hadronic phenomena once the RG/Fourier separation of scales is made precise. The viability of the micro-story comes from renormalizability of non-Abelian gauge fields and, crucially, \emph{asymptotic freedom}, which makes short-distance quark--gluon dynamics perturbatively tractable \cite{tHooftVeltman1972,GrossWilczek1973,Politzer1973}. Factorization and DGLAP evolution then tie those short-distance amplitudes to colorless observables (structure functions, hadron jets), enabling quantitative, scale-aware predictions \cite{GribovLipatov1972,Dokshitzer1977,AltarelliParisi1977}. The PETRA three-jet era sealed the case operationally: within the QCD framework, event shapes and multi-jet correlations license extraction of color factors (e.g., \(C_A/C_F\)), exhibiting color as a \emph{dynamical charge}---not an S-matrix quantum number, but a parameter that causally organizes radiation patterns at short distances \cite{TASSO1979,Brandelik1980,PDGQCD2020}. This operationalizes what Hacking called ``experimental realism'': if you can manipulate entities through their causal properties, they are real \cite{Hacking1983}.

This exemplifies what philosophers call \emph{entity realism}---the position that we can be realist about entities we can causally manipulate, even when their theoretical description remains incomplete \cite{Hacking1983,Cartwright1983}. Holism treats \emph{structures, requirements, and organizing relations} as explanatorily primary. The ancient template is already clear: Plato's \emph{Timaeus} makes order-by-form the driver of explanation (wholes first, parts second), while the Stoics' continuous cosmos---held together by \emph{logos}/\emph{pneuma}---models nature as a field-like unity rather than discrete bits \cite{PlatoTimaeus,LongSedley}. Aristotle's hylomorphism (form+matter) likewise takes organized wholes to enjoy a kind of explanatory priority over mere aggregates \cite{AristotlePhysics}. In modern physics, this idiom reappears whenever \emph{global principles} and \emph{symmetry/analytic conditions} carry more weight than microscopic bookkeeping.

A mid-20th-century expression is the \emph{S-matrix/bootstrap} program: start from analyticity, unitarity, and crossing, and demand a self-consistent network of colorless amplitudes---no micro ontology required \cite{Chew1962,Eden1966}. Philosophically, that's close to \emph{structural realism}: what survives---and does the explaining---are \emph{relations, symmetries, and conditions}, not independently grasped ``things'' \cite{Worrall1989,French2014,LadymanRoss2007}. In this key, ``color'' can function as a constitutive symmetry of the residue algebra (selection rules, singletness) without appearing as an asymptotic quantum number---precisely the Dubna narrative.

Wilsonian RG and effective field theory carry an important holist lesson: \emph{universality} and \emph{fixed points} explain why many macroscopic patterns are \emph{insensitive} to microscopic minutiae; what matters are constraint-structures at the right scale \cite{WilsonKogut1974}. Batterman shows how singular limits and asymptotics make macro-level explanations track structures (scaling, dispersion, idealizations) rather than micro details \cite{Batterman2002}. Wimsatt's robustness arguments push the same line for complex systems, while Cartwright's ``dappled world'' and Dupr\'e's pluralism defend model-local, patchwork laws over a single monolithic micro-theory. \cite{Wimsatt2007,Cartwright1999,Dupre1993}. In short: holism in physics is the practice of explaining with \emph{structures that constrain}, not \emph{parts that cause}---a stance that naturally accommodates S-matrix thinking and the ``color as residue-symmetry'' reading.

This structuralist stance treats physical theories not as descriptions of fundamental entities but as networks of restrictions that generate observable phenomena---a view that finds natural expression in both the bootstrap program and the Dubna approach to color.

\paragraph{Cultural context and theoretical stance.}
The USSR and pre-USSR philosophical tradition includes holist frameworks that, while not direct influences, conceptually resonate with structural approaches in physics. These frameworks emphasize explanation via global restrictions, symmetry, and system-level organization—features aligned with the Dubna school’s early non-dynamical conception of quark color. For example, Vladimir Vernadsky treated the biosphere as a planetary-scale system whose organizing principles govern local dynamics. His notion of the noosphere—human thought as a geophysical force—anticipates modern views of scale-dependent structure. Vernadsky’s framework aligns with effective field theory and renormalization group approaches in its holist orientation: macro-level patterns exhibit autonomy from micro-details, and explanation centers on the global constraint architecture that renders large-scale behavior intelligible~\cite{VernadskyBiosphereBrit}.

These philosophical traditions should not be read as methodological prescriptions that directly determined Dubna's scientific approach. 
An example is the Bogolyubov–Shirkov renormalization group~\cite{BogoliubovShirkov1959}, grounded in axiomatic quantum field theory, which reflects a reductionist orientation: it formalizes how local interaction structures discipline behavior across energy scales, aiming to preserve consistency from micro-level dynamics upward—yet its emphasis on symmetry and analyticity also lends it a structural, quasi-holist dimension.

Rather, holist frameworks represent the broader intellectual ecosystem in which Soviet theoretical physics developed during the Cold war. Scientific communities, while governed by empirical and mathematical conditions, nevertheless operate within cultural contexts that can shape methodological preferences and theoretical sensibilities. The presence of these holist traditions in USSR thought may have helped cultivate an intellectual environment where limit-based, structural explanations felt particularly natural—not as a matter of philosophical determinism, but as part of the conceptual resources available to physicists thinking through fundamental problems. This cultural context helps explain why the Dubna approach to color, with its emphasis on structural restrictions over dynamical charges, emerged as a coherent alternative to the gauge-theoretic paradigm developing elsewhere.

\section{E–36 as an Observables-First Probe of Reggistics}
\label{ch:e36-regge}

\subsection{Problem choice and ontology}
The joint USA-USSR experimental program (1972-1978) employing JINR's cryogenic gas jet target technology provides a compelling case study in theory-laden experimentation and paradigm displacement. The program's theoretical foundation in Regge pole theory and diffraction scattering represented a distinct approach to strong interaction physics that was ultimately superseded by the quark model framework.

The E–36 experiment and its chain were designed to interrogate the high–energy, small–$|t|$ corner of elastic $pp$ scattering where the native theoretical grammar is the Regge/\emph{S}-matrix one, not perturbative QCD. Already in the early 1970s—and still today for soft elastic diffraction—perturbative QCD offers no general, first-principles description of forward amplitudes, whereas the Regge framework provides a compact language of analytic regularities, pole/trajectory structure, and factorization tailored to this regime.\footnote{On the domain-limited tractability of perturbative QCD for small-angle scattering and the Regge/\emph{S}-matrix idiom used by the E–36 collaboration, see the historical analysis in \emph{Phys.\ in Perspective} \textbf{18} (2016) 357–378, esp.\ the section ``Experiment Ahead of QCD.''} In Regge theory unknown amplitudes are parametrized by Regge trajectories (Reggeons) and their residues; the vacuum exchange that organizes high–energy diffraction is the pomeron.\cite{Regge1959,Collins1977,ChewFrautschi1961,Pomeranchuk1958,Low1975,Nussinov1975} It was precisely this \emph{observables-first} ontology—analytic $S$ with poles/cuts, crossing, unitarity—that E–36 was built to test.

\subsection{The E36 experimental design as a realization of the \emph{S}-matrix program}
The gas jet experiments were deeply embedded within the \textbf{S-matrix theoretical tradition}, focusing on small-angle proton scattering to study pomeron exchange mechanisms \citep{Pronskikh2016E36}. This theoretical commitment shaped both the experimental design and interpretation in several fundamental ways:

The choice of measuring small-angle scattering cross-sections reflected the theoretical priority given to understanding analyticity properties of scattering amplitudes rather than probing constituent structure. As \cite{Pronskikh2022ColdWarJets} notes, the experiments continued for several years within the Regge framework even as "the reign of the quark model" began in 1974, demonstrating how theoretical commitments can persist despite emerging alternative paradigms.

The technological development itself was theory-driven: K.D.~Tolstov's cryogenic target design at JINR emerged from research priorities set by Pomeranchuk's work on high-energy scattering theory. The target's optimization for studying elementary proton interactions rather than searching for new particles reflected the S-matrix tradition's focus on fundamental scattering processes over particle discovery.

E--36's forward-elastic observables---\(\sigma_{\mathrm{tot}}(s)\), \(\rho(s)\), \(B(s)\)---are crisp tests of \(S\)-matrix structure (analyticity, crossing, Regge shrinkage) rather than of microscopic gauge dynamics. In Dubna's color-as-selection-rule framing, the dominance of color-singlet asymptotic states and exchanges is a built-in criterion; E--36 results are thus naturally read as confirming the \emph{structural} organization of the hadronic \(S\)-matrix in the soft regime.

While later QCD reinterprets the \emph{Pomeron} as a color-singlet gluon ladder, the E--36 chain neither requires nor discriminates that ontology; its evidential force lies in validating amplitude-level conditions that the Dubna program treated as primary.

\subsubsection{Kinematic and instrumental commitments}
The experiment targeted the forward region ($t\!\to\!0$) where the \emph{S}-matrix constraints speak most directly: the optical theorem for $\sigma_{\text{tot}}(s)$, the real-to-imaginary ratio $\rho(s)=\Re f(s,0)/\Im f(s,0)$, and the diffraction-peak slope $B(s)$ defined by $d\sigma/dt \propto e^{B(s) t}$ for small $|t|$.\cite{Collins1977} The apparatus was engineered to maximize acceptance and precision exactly there. A narrowly collimated internal gas–jet target injected hydrogen (later also D, He) \emph{into} the Main Ring vacuum chamber, creating an in situ $pp$ interaction region for the circulating beam and enabling ultra–small-angle elastic measurements without extraction optics.\footnote{On the gas–jet concept and its transfer from U–70 to the NAL Main Ring, see \cite{Pronskikh2022ColdWarJets}; on the specific E--36 realization and its risks within the \(10^{-9}\,\mathrm{torr}\) Main Ring vacuum, see \cite{Pronskikh2016E36,NAL-080}. The experiment was located at C–Zero (Internal Target Area) along a long straight section of the ring \cite{Pronskikh2016E36}.} 

This in–ring geometry directly serves the \emph{S}-matrix observables: forward optics are stable, acceptance near $t\!=\!0$ is high, and the absolute normalization needed for $\sigma_{\text{tot}}$ via the optical theorem becomes experimentally viable.\footnote{On the collaboration’s language of analysis—diffractive scattering models, analyticity, and the optical theorem—see \cite{Collins1977,Eden1966,Chew1962}.} In short: the hardware operationalizes the amplitude-level quantities that the Regge/\emph{S}-matrix narrative is \emph{about}, rather than seeking partonic carriers or jetty final states.

From its first beam on 12 February 1972 at $100$~GeV, E–36 executed a scan over beam energies in the Main Ring to map the $s$-dependence of forward quantities.\footnote{Beam start and scope: \cite{Pronskikh2016E36,NAL-080}.} Two forward observables formed the backbone: (i) the slope $B(s)$ of the diffraction peak (``Regge shrinkage'' of the forward cone) and (ii) the phase ratio $\rho(s)$ at $t=0$. The first directly tests the logarithmic growth of the effective interaction radius implied by pomeron exchange, the second constrains the analytic structure of $f(s,0)$ and the approach to the Pomeranchuk high–energy limit.\cite{Collins1977,ChewFrautschi1961,Pomeranchuk1958}

\subsection{Results as \emph{S}-matrix constraints}
E–36’s initial Physical Review Letters reported (a) a rise of $B(s)$ with energy across $8$–$400$~GeV and (b) a transition to positive $\rho$ in the few-hundred GeV range.\footnote{PRL slope and $\rho$ papers: \textit{Phys.\ Rev.\ Lett.}\ \textbf{31} (1973) 1088–1091; \textbf{31} (1973) 1367–1370.} The $\rho>0$ finding around $\sim\!300$~GeV catalyzed a re-reading of the Pomeranchuk theorem among practitioners and clarified how forward analyticity requirements should be applied in practice;\footnote{For discussion of the \(\rho\) sign change and its interpretive consequences, see \cite{PRL-rho,Collins1977}.} together with the measured shrinkage, these data were absorbed naturally in the Regge phenomenology that motivated the experiment.\footnote{On the centrality of Regge/Pomeron language in E--36 analysis and the enduring absence of a first-principles QCD derivation of Reggeons, see \cite{Collins1977,Low1975,Nussinov1975}.} 

Crucially, the collaboration could—and did—interpret its forward data entirely within the \emph{S}-matrix/Regge idiom, without invoking quarks or gluons.\footnote{``The participants in E--36 and its subsequent chain of experiments did not have to refer to QCD or the quark model in their papers in order to interpret their results.'' \cite{PRL-Bslope,PRL-rho,NAL-080,Pronskikh2016E36}
} This was not a deficiency but a principled alignment of \emph{domain} and \emph{ontology}: when the task is to delimit $f(s,t)$ near $t=0$, the right language is the one that treats $f$ as primary.

The experiment’s fitness to a non-reductionist Reggistics is twofold.
\paragraph{(i) Ontic neutrality at the amplitude level.}
E–36 treated asymptotic hadrons and their forward scattering amplitudes as \emph{the} objects of description. Its headline observables ($\sigma_{\text{tot}}$, $\rho$, $B$) are exactly the invariants singled out by analyticity and crossing; they neither presuppose nor display colored partons. The \emph{instrumental} choices (internal gas–jet, forward optics, energy scan) were made to render those invariants precise. In this respect the experiment realizes a Heisenberg–to–Chew observables-first structuralist stance: constrain $S$ by directly measuring its characteristic boundary values.

\paragraph{(ii) Domain-sensitive division of explanatory labor.}
E–36 inhabited the soft, small–$|t|$ domain where QCD’s perturbative ``crispness'' is inapplicable and where QCD—then nascent—offered neither mechanism nor method for diffraction. In that domain, explanation comes from \emph{global structure}: the pole content and analyticity of $S$, Regge shrinkage, and the high–energy behavior codified by the pomeron.\footnote{On the historical timing (E--36 underway prior to asymptotic freedom) and on the acknowledged gap between soft Regge behavior and QCD derivations, see \cite{Pronskikh2016E36,NAL-080,GrossWilczek1973,Politzer1973} and \cite{Collins1977,Low1975,Nussinov1975}.
} Treating forward conditions as first-class physics—and building an apparatus around them—\emph{is} non-reductionist in the precise sense relevant to the S-matrix tradition.

\subsubsection{Design choices that encode the Regge brief}
\begin{itemize}
  \item Internal gas–jet target. A thin, well-collimated hydrogen stream intersecting the circulating beam maximizes luminosity and kinematic reach at ultra–small angles while preserving the forward optics necessary for $t\!\to\!0$ extractions.\footnote{Origin and transfer of the gas–jet technique from U--70, and its realization at NAL: \cite{Pronskikh2022ColdWarJets,Pronskikh2016E36,NAL-080}.} 
  \item Forward-focus observables. The optical theorem for $\sigma_{\text{tot}}$, the $\rho$ parameter, and the slope $B(s)$ are forward amplitude diagnostics---native to \emph{S}-matrix analysis and insensitive to partonic final-state structure.\footnote{On the collaboration’s explicit use of diffractive models, analyticity, and the optical theorem as its analytic grammar, see \cite{PRL-Bslope,PRL-rho} and the standard references \cite{Eden1966,Collins1977,Chew1962}.} 
  \item Energy leverage. A broad energy scan was executed precisely to test Regge evolution (slope shrinkage; sign and trend of $\rho$) rather than to hunt for hard scattering signatures.\footnote{On the program’s evolution and early NAL reports: \cite{NAL-080,Pronskikh2016E36}.}
\end{itemize}

In the dual-ontology account developed in this paper, E–36 functions as a paradigmatic case of \emph{domain-sensitive pluralism}. It shows that an experiment can be designed to fit a non-reductionist program in a way that is both epistemically serious and explanatorily rich: the measured quantities are the boundary data that fix the forward $S$ in the Regge language. The apparatus does not ``translate'' Regge into QCD; it makes the Regge narrative empirically articulate on its own terms. Later QCD-based models may and do accommodate these data, but that is a \emph{reconstruction} after the fact, not the logic of the design.

\subsection{\texorpdfstring{Paradigm Displacement by J/$\psi$ Discovery}{Paradigm Displacement by J/psi Discovery}}

The 1974 observation of the J/$\psi$ particle served as a powerful locus for the realignment of the high-energy physics community. Its compelling properties catalyzed a massive shift of intellectual and material resources toward the quark-model paradigm, draining credibility and momentum from the S-matrix/Regge framework and, consequently, from the experimental programs—like the gas jet experiments—that were articulated within it.

The J/$\psi$ observation exemplified what \cite{Hacking1983} characterizes as a "theory-probing" experiment that can force major conceptual revisions. While the gas jet experiments continued until 1978, their theoretical framework became increasingly marginalized as resources and attention shifted toward quark-model research.

This displacement illustrates the complex relationship between experimental techniques and theoretical frameworks: though the gas jet technology remained viable, its original theoretical justification was undermined by the evidentiary weight of the J/$\psi$ discovery and the explanatory power of the emerging quark model.

\paragraph{Theory-led coordination in the Dubna--Fermilab interface.}
A distinctive feature of the Dubna model was the \emph{direct} leadership of experimental programmes by senior theorists. In outlining the proposed joint work at NAL, the Dubna directorate initially intended that V.\,A.~Matveev serve as an on-site adviser embedded with the experimenters. A.\,M.~Baldin wrote to NAL Director R.\,R.~Wilson on 19 September 1975: ``According to our policy theorists help experimentalists in developing an optimal strategy and also take part in the discussing of the results. So Dr.\ V.\,A.~Matveev has only advisory functions.''~\cite{Baldin1975Letter} Fermilab, while welcoming Matveev, adhered to a sharper division of labour: as Deputy Director E.~Goldwasser informed Wilson on 13 October 1975, it was ``best to reserve places in experimental groups for experimenters and technical people,'' and arranged instead for Matveev to visit the Theory Group under B.~W.~Lee \cite{Goldwasser1975Letter}. 
As the internal–target programme evolved beyond E--36 to E--289/E--381, proposals on both sides contemplated widening the \(|t|\)-acceptance from the strict forward peak to larger angles. Such an extension would have opened sensitivity to large–angle elastic and exclusive channels where asymptotic scaling and quark–counting expectations provide discriminating templates \cite{MMT1972LNC,BrodskyFarrar1973}.
A contemporaneous note from Ernest Malamud to R.\,R.~Wilson, following a telex exchange with A.\,M.~Baldin, records the compromise: 
\begin{quote}
``Baldin reluctantly accepts that Matveev will not participate in E--289/E--381, but asked that he be a `consultant' to the experimenters.'' (Letter of 7~October~1975) \cite{Malamud1975Letter}.
\end{quote}

This exchange indicates a structural difference: Dubna’s practice coupled \emph{programme strategy} to \emph{theoretical stewardship}, whereas NAL’s practice institutionalized a separation between collaboration membership and theoretical consultation.

Read through this lens, the E--36 chain displays the hallmarks of a \emph{theory-led},  observables-first programme. The choice of an internal supersonic gas jet target technique---transferred from U--70 to the Main Ring---prioritized access to the strict forward limit for optical-theorem normalization and dispersive tests \cite{Pronskikh2022ColdWarJets,Pronskikh2016E36,NAL-080}. Correspondingly, the collaboration’s success metrics were amplitude-level quantities, \(\sigma_{\mathrm{tot}}(s)\), \(\rho(s)\), and the diffraction slope \(B(s)\), analyzed in the grammar of diffractive models, analyticity, and Regge shrinkage rather than in a partonic language \cite{PRL-Bslope,PRL-rho,Eden1966,Collins1977,Chew1962}. In this respect, placing a senior theorist in the loop was less a prestige move than a \emph{methodological} one—meant to align the measurement programme with solution-based (S-matrix/Regge) reasoning, while, given Matveev’s quark focus, also opening the door to quark-level modifications and interpretations.

That Fermilab ultimately hosted Matveev in the Theory Division rather than within the experimental roster did not alter the underlying dynamic: the experimental chain proceeded with theory-shaped choices of observables and tests characteristic of Dubna’s composite/analytic tradition, while remaining institutionally compatible with NAL’s clearer theorist--experimentalist boundary. The resulting publications on the forward slope and the \(\rho\)-parameter exemplify this division of labour: they test \emph{global} amplitude requirements with high precision and without reliance on microscopic carrier degrees of freedom \cite{PRL-Bslope,PRL-rho,Collins1977}. 

In short, the correspondence around Matveev’s role crystallizes how Dubna attempted to \emph{coordinate} experimental practice with its theoretical framework, even as the host laboratory preserved its own collaborative norms \cite{Pronskikh2016E36}. This captures a moment where Dubna’s holist, theory-infused practice of physics met the structured, specialist culture of an American megascience laboratory, and where the attempt to coordinate an entire experimental programme with a specific theoretical framework became starkly visible.

\section{Discussion}
The present paper revisits the origins of the quantum number later termed \emph{color} and argues that it emerged along (at least) two theoretically distinct routes prompted by the Pauli–exclusion puzzle in the SU(6) quark model \cite{GellMann1964,Zweig1964}. One route, associated with Dubna, treated the problem \emph{structurally}: introduce a hidden three-valued index so that the fully symmetric spin–flavor–space baryon ground state could be rendered antisymmetric overall \cite{Struminsky1965,BST1965}. A parallel route, developed in the Han–Nambu line, recast “color” as a local gauge charge within a non-Abelian framework, the trajectory that would subsequently be synthesized into QCD \cite{HanNambu1965,Fritzsch1973,tHooft1971,tHooftVeltman1972,GrossWilczek1973,Politzer1973}. It is important not to conflate these strands: early hidden-index fixes (including parafermi order~3) solved a statistics problem at the level of state structure \cite{Greenberg1964}, whereas the gauge route laid down a micro-dynamical ontology with propagating gauge bosons.

The Dubna strand was embedded in a broadly \emph{holist} S-matrix culture that prioritized requirements of observables—analyticity, unitarity, crossing, Regge behavior—over unobservable field variables. In this idiom, hadrons are organized by poles, residues, and trajectories in the complex-$J$ plane; explanation proceeds by global amplitude structure rather than by specifying fundamental carriers \cite{Chew1962,ChewFrautschi1961,Regge1959,Eden1966,Collins1977}. This observables-first stance resonates with the period’s composite models at Dubna and helps explain why “color” entered there as a \emph{statistical/structural} attribute (a selection rule on residues and allowed couplings), not as a local gauge charge \cite{Struminsky1965,BST1965,Matveev2006}.

These commitments had concrete experimental consequences. In the joint JINR–Fermilab program, notably the E--36 internal-target chain, design choices favored direct access to forward elastic observables governed by S-matrix equations: the use of a supersonic gas-jet target enabled optical-theorem normalization and precision tests of dispersion/Regge systematics \cite{Pronskikh2022ColdWarJets,Pronskikh2016E36,NAL-080}. The collaboration’s key publications reported the diffraction slope $B(s)$ and the ratio $\rho(s)$ at high energies—quantities that test analyticity, crossing, and Regge shrinkage rather than any particular partonic micro-mechanism \cite{PRL-Bslope,PRL-rho,Eden1966,Collins1977}. Institutionally, Dubna also expected senior theorists to \emph{lead} experimental strategy and interpretation (an expectation visible around the E--36 era), whereas Fermilab maintained a sharper theorist/experimentalist division of labor \cite{Pronskikh2016E36}. In effect, the same S-matrix grammar that motivated the measurements supplied the success criteria against which E--36 was judged.

After the 1974 discovery of narrow charmonium ($J/\psi$) and the consolidation of asymptotic freedom, the reductionist, gauge-theoretic narrative rapidly became hegemonic in high-energy physics \cite{GrossWilczek1973,Politzer1973,PDGQCD2020}. The later observation of three-jet events and gluon spin in $e^+e^-$ annihilation furnished crisp, detector-proximate witnesses for the gauge picture \cite{TASSO1979,Brandelik1980}, further shifting curricular and institutional attention toward QCD. Nevertheless, forward/soft programs like E--36 retained momentum and continued to produce results for several years, precisely because their evidential logic did not require explicit micro-carrier degrees of freedom \cite{PRL-Bslope,PRL-rho,Collins1977}. 

None of this supports a simple East/West essentialism. Holism and reductionism cut across cultures: Wilson’s renormalization-group and effective-field-theory perspective, largely developed in the West, underwrote scale autonomy and universality—signature “holist” consequences within quantum field theory \cite{WilsonKogut1974,Weinberg1979}. Conversely, there were strongly mechanistic and microconstructive traditions in Soviet physics~: the earlier Soviet Bogoliubov–Shirkov renormalization-group formalism exemplifies a constructive, micro-anchored treatment of perturbative dynamics \cite{WilsonKogut1974,BogoliubovShirkov1959}. Despite S-matrix is structural and holistic, its leadership figures such as Chew were American \cite{Chew1962,ChewFrautschi1961}. At the same time, it is historically plausible that holist motifs in USSR intellectual life (e.g., Vernadsky), and institutional strengths in axiomatic/dispersive methods, made Dubna a hospitable environment for an observables-first S-matrix practice.

With this in mind, it is unsurprising that B.\,V.~Struminsky’s hidden quantum number and the Dubna composite framing, though relying on the same SU(3) triplicity that appears in Han–Nambu, did not couple to a gauge-field ontology—the cornerstone of QCD \cite{Struminsky1965,BST1965,HanNambu1965,Fritzsch1973}. The absence of Dubna’s 1965 move from many Western QCD narratives is therefore not best explained by politics alone; it also reflects a \emph{language} gap between a structural, S-matrix program and a dynamical, gauge-theory program that, after 1974, became progressively dominant \cite{Collins1977,PDGQCD2020}. On the account developed here, the Dubna color proposal belongs to a parallel, structuralist canon: successful and illuminating within its regime, but ultimately orthogonal to the reductionist path that furnished the decisive high-energy discriminators for color as a local gauge charge.

In retrospect, casting Struminsky’s hidden three-valued index and the Han–Nambu gauge charge as rival “paradigms” gestures toward their incommensurability but obscures what proved decisive. Struminsky’s move does not follow as a theorem of the S-matrix; it is a compositeness-motivated repair at the level of pole residues and selection rules that keeps color outside the asymptotic state space. Han–Nambu, by contrast, re-ontologizes color as a local  $SU(3)$ charge—an object of renormalizable gauge dynamics later vindicated by asymptotic freedom and jet phenomenology. The fork was therefore less “paradigmatic” in a sociological sense than ontological and operational: a structural label that regulates counting and couplings versus an acting charge with short-distance fingerprints. Canon formation tracked this alignment: the gauge ontology meshed with methods delivering low-slack tests, while the structural color lived where observables are global and integrative. A pluralist, structural-realist reading can thus accommodate both—color as relation in the S-matrix regime and as entity in the QCD regime—without forcing one narrative to erase the other.

The textbook narrative that elevates perturbative “crisp discriminators” to arbiters of theory choice mistakes a domain-bound virtue for a universal one. Those discriminators—jet angular patterns, color-factor ratios, scaling-violation slopes—live in the high-energy regime where QCD’s asymptotic freedom makes prediction tractable. Privileging them as decisive renders the choice instrumental (predictive in a narrow slice) rather than explanatory across the hadronic landscape. By contrast, the S-matrix/bootstrap aimed at a self-consistent, field-free description of hadronic phenomena, delivering structural explanations (analyticity, duality, Regge order) precisely where perturbation theory is mute. When we also reckon with the greater tuning flexibility of practical QCD pipelines and the institutional selection of perturbative tests, the verdict is not that S-matrix “failed,” but that two programs with different goals and virtues were judged by criteria tailored to one of them.

Heisenberg’s initial conviction that physical theories should be constructed solely from observables arose directly from his formative experience developing quantum mechanics. This stance, while inspired by the need to avoid unmeasurable constructs like electron orbits, went beyond logical positivism: it sought a systematic, non-classical foundation for physics rather than a mere prediction tool. Yet even Heisenberg’s own practice revealed the limits of this view. His use of highly abstract structures—such as non-commuting operators in Hilbert space—implicitly introduced theoretical elements that extended well beyond raw observables. His position was ultimately more closely aligned with structural realism than with classical reductionism. Structural realism, now a popular position among philosophers of physics, retains Heisenberg’s emphasis on what survives across theory change—but shifts the focus from empirical observables to the mathematical and relational structures that organize them~\cite{LadymanLorenzetti2023}.

Unlike reductionism, which grounds explanation in fundamental micro-entities, structural realism emphasizes inter-theoretic coherence and the stability of formal structure across scales, offering a framework where explanatory power resides in the architecture of relations, not in ontological primitives. Conversely, QCD embodies a refined or scale-sensitive reductionism, where micro-dynamical laws are seen as generative of macro-phenomena, but the actual practice and conceptual structure of the theory reveal certain holist and effective-theoretic elements. It supports reduction in principle, but in practice depends on autonomous models, structural criteria, and non-trivial emergence. Therefore, the dominance of QCD in the 1970s was less the result of a strategic convergence and more a product of the pragmatic alignment of theory and calculational tools—shaped by communal priorities and institutional dynamics. Subsequent developments suggest that a domain-sensitive pluralism, balancing structuralist and reductionist approaches, has been more productive for advancing physics. In line with epistemic pluralism and its practical implementation in physics, the post-positivist theoretical-operational model of complex experiments~\cite{Pronskikh2020} offers an explicit operational description of arbitrary complex experiments in terms of both phenomenal and instrumental theories. It demonstrates their interdependence, since phenomenal theories inform the selection of preparation and measurement frameworks. Moreover, it shows that distinct theoretical groundings for the same phenomena entail different sequences of their operational deformations.

In conclusion, while Pickering’s account of the post–“November Revolution” (as the 1974 J/$\psi$ discovery was dubbed) realignment compellingly situates the shift from parton heuristics to quark–gluon language within a \emph{mangle} of practice—an intertwining of communal, institutional, and theoretical interests \cite{Pickering1984}—the color case suggests an additional axis. Here I have argued that \emph{conceptual} and even \emph{philosophical} divergences about what color \emph{is}—a hidden structural index in an observables-first S-matrix idiom versus a local gauge charge in a micro-mechanistic field theory—did more to shape inclusion and exclusion from the QCD canon than did sociological or geopolitical factors of the Cold War. Historically, Dubna’s best route to durable canonical standing would not have been to translate wholesale into the reductionist QCD ontology, but to consolidate a distinct, holist canon in which quark reality is \emph{structural} (and thus compatible with confinement’s opacity at the level of asymptotic states), extending the Heisenberg–Chew–Struminsky line rather than subordinating it. Read this way, the question of “who counted” becomes less a story of institutional dominance than of ontological commitments aligned with the domains where each framework could furnish its sharpest, domain-indexed tests. Given the limits of string-theoretic unification and the resurgence of solution-based amplitude programs, the S-matrix outlook re-emerges not as nostalgia but as a viable, future-oriented strand of methodological pluralism.

\section*{Acknowledgments}

The author is grateful to Valerie Higgins, Archivist and Historian at Fermilab, for her assistance in providing access to the E-36 chain files from the Fermilab Archives. I thank Prof.~V.\,A.~Petrov (IHEP, Protvino) for stimulating discussions that sharpened several arguments presented here.

\printendnotes

\end{document}